\newcommand{\GCD}{{\rm{GCD}}}

\newcommand{\st}{{\;|\;}}
\newcommand{\pn}{{\noindent}}

\newcommand{\lbrk}{{\linebreak[0]}}		
\newcommand{\realtilde}{\makebox{\tt\char'176}}
\newcommand{\beq}{\begin{equation}}
\newcommand{\eeq}{\end{equation}}
\newcommand{\calB}{\makebox{$\cal B$}}
\newcommand{\calD}{\makebox{$\cal D$}}

\newcommand{\calH}{\makebox{$\cal H$}}
\newcommand{\calI}{\makebox{$\cal I$}}

\newcommand{\calU}{\makebox{$\cal U$}}
\newcommand{\calV}{\makebox{$\cal V$}}
\newcommand{\KnuthG}{\makebox{$\widetilde{F}$}}

\newcommand{\Val}{{\rm{Val}}}
\newcommand{\tcU}{{\cal{\widetilde{U}}}}

\newcommand{\lla}{\makebox{\hbox to 30pt {\leftarrowfill}}}
\newcommand{\lra}{\makebox{\hbox to 30pt {\rightarrowfill}}}

\newcommand{\QED}{{\hspace*{\fill}\hbox{\rlap{$\sqcap$}$\sqcup$}}} 

\newcommand{\ia}{\hspace*{1em}}			
\newcommand{\ib}{\hspace*{2em}}

\documentclass[11pt]{article}
\newtheorem{lemma}{Lemma}

\newtheorem{thm}{Theorem}

\begin{document}
\bibliographystyle{plain}

\begin{center}
{\bf \large Further analysis of the Binary Euclidean algorithm}\\[10pt]
Richard P. Brent%
\footnote{
\hbox{Copyright \copyright 1999, 2013 R.~P.~Brent.
Present address: MSI, ANU, ACT 0200, Australia} \hspace*{\fill}
\hbox{rpb183tr}}\\
Oxford University\\[5pt]
Technical Report PRG TR-7-99\\[5pt]
4 November 1999
\end{center}

\begin{abstract}

The binary Euclidean algorithm is a variant of the
classical Euclidean algorithm.
It avoids multiplications and divisions, except by powers of two,
so is potentially faster than the classical algorithm
on a binary machine.
\medskip

We describe the binary algorithm and consider its average case behaviour.
In particular, we correct some errors in the literature,
discuss some recent results of Vall\'ee, and describe a numerical
computation which supports a conjecture of Vall\'ee. 

\end{abstract}

\section{Introduction}
\thispagestyle{empty}

In \S\ref{sec:bineuc} we define the {\em binary} Euclidean algorithm
and mention some of its properties, history and generalisations.
Then, in \S\ref{sec:heuristic} we outline the heuristic model
which was first presented in 1976~\cite{rpb037}.
Some of the results of that paper are mentioned (and simplified)
in \S\ref{sec:empirical}.
\medskip

Average case analysis of the binary Euclidean algorithm lay dormant
from 1976 until Brigitte Vall\'ee's recent analysis~\cite{Va-ANTS,Va-Algo}.
In \S\S\ref{sec:algV}--\ref{sec:vallee} we discuss Vall\'ee's results and
conjectures. In \S\ref{sec:numerical} we give some numerical evidence
for one of her conjectures.
Some connections between Vall\'ee's results and our earlier
results are given in \S\ref{sec:formal}.
\medskip

Finally, in \S\ref{sec:correction} we take the opportunity to point
out an error in the 1976 paper~\cite{rpb037}. Although the error is
theoretically
significant and (when pointed out) rather obvious, it appears that
no one noticed it for about twenty years. The manner of its discovery
is discussed in \S\ref{sec:correction}.
Some open problems are mentioned in~\S\ref{sec:conc}.

\subsection{Notation}

$\lg(x)$ denotes $\log_2(x)$. 
$N, n, a, k, u, v$ are positive integers.
\medskip

\noindent $\Val_2(u)$ denotes the dyadic valuation of the positive integer $u$,
i.e.\ the greatest integer $j$ such that $2^j \;|\; u$.
This 
is just the number of
trailing zero bits in the binary representation of $u$.
\medskip

$f, g, F, \KnuthG, G$ are functions of a real or complex variable,
and usually $f(x) = F'(x)$, $g(x) = G'(x)$ etc.
Often $f, g$ are probability densities and $F, G$ are the
corresponding probability distributions.
\medskip

\noindent {\bf Warning}:
Brent~\cite{rpb037},
Knuth~\cite{Knuth97},
and Vall\'ee~\cite{Va1,Va-ANTS,Va-Algo}
use incompatible notation.
Knuth uses $G(x)$ for our $\KnuthG(x)$, and $S(x)$ for our $G(x)$.
Vall\'ee sometimes interchanges our $f$ and $g$.

\vspace*{\fill}\pagebreak[4]
\section{The Binary Euclidean Algorithm}
\label{sec:bineuc}

The idea of the {\em binary} Euclidean algorithm is to avoid
the ``division'' operation $r \leftarrow m \bmod n$ of the classical
algorithm, but retain $O(\log N)$ worst (and average) case.
\medskip

We assume that the algorithm is implemented on a binary computer so
division by a power of two is easy. In particular, we assume that
the ``shift right until odd'' operation
\[ u \leftarrow u/{2^{\Val_2(u)}} \]
or equivalently
\medskip

\begin{center}
while even($u$) do $u \leftarrow u/2$
\end{center}
\medskip

\noindent can be performed in constant time, although time $O(\Val_2(u))$
would be sufficient. 

\subsection{Definitions of the Binary Euclidean Algorithm}

There are several almost equivalent ways to define the algorithm.
It is easy to take account of the largest power of two dividing the inputs,
using the relation
\[\GCD(u,v) = 2^{\min(\Val_2(u),\Val_2(v))}\;
	\GCD\left(u/2^{\Val_2(u)},v/2^{\Val_2(v)}\right)\;,\]
so for simplicity we assume that $u$ and $v$ are {\em odd}
positive integers.
Following is a simplified version of the algorithm given
in Knuth~\cite[\S4.5.2]{Knuth97}.

\begin{description}
\item[Algorithm B] 
\item[B1.] $t \leftarrow |u-v|$;\\
	if $t = 0$ terminate with result $u$
\item[B2.] $t \leftarrow t/{2^{\Val_2(t)}}$
\item[B3.] if $u \ge v$ then $u \leftarrow t$ else $v \leftarrow t$;\\
	go to B1.
\end{description}

\subsection{History}

The binary Euclidean algorithm is usually attributed to
Silver and Terzian~\cite{Silver}
or (independently) Stein~\cite{Stein} in the early 1960s. 
However, it seems to go back much further.
Knuth~\cite[\S4.5.2]{Knuth97} 
quotes a translation of a first-century AD Chinese text
{\em Chiu Chang Suan Shu} on how to reduce a fraction to lowest terms:

\begin{quote}
If halving is possible, take half.\\[5pt]
Otherwise write down the denominator and the numerator,\\
and subtract the
smaller from the greater.\\[5pt]
Repeat until both numbers are equal.\\[5pt]
Simplify with this common value.
\end{quote}

\noindent This is essentially Algorithm~B.
Hence, the binary algorithm is almost as old as the classical
Euclidean algorithm~\cite{Euclid}.

\subsection{The Worst Case}

Although this paper is mainly concerned with the average case behaviour
of the binary Euclidean algorithm, we mention the worst case briefly.
At step B1, $u$ and $v$ are odd, so $t$ is even. Thus, step B2 always
reduces $t$ by at least a factor of two. Using this fact, it is easy to
show that $\lg(u+v)$ decreases by at least one each time step B3 is executed,
so this occurs at most
\[ \lfloor\lg(u+v)\rfloor \]
times~\cite[exercise 4.5.2.37]{Knuth97}.
Thus, if $N = \max(u,v)$,
step B3 is executed at most
\[ \lg(N) + O(1) \]
times.
\medskip

Even if step B2 is replaced by single-bit shifts
\medskip
\begin{center}
while even($t$) do $t \leftarrow t/2$
\end{center}
\medskip
the overall worst case time is still $O(\log N)$. 
In fact, it is easy to see that $(\lg(u) + \lg(v))$ decreases by at least
one for each right shift, so the number of right shifts
is at most $2\lg(N)$.

\subsection{The Extended Binary Algorithm}

It is possible to give an {\em extended} binary GCD algorithm
which computes integer multipliers $\alpha$ and $\beta$ such that
\[ \alpha u + \beta v = \GCD(u,v)\;. \]
Let $n = \lceil \lg u \rceil + \lceil \lg v \rceil$ be the number of
bits in the input.
Purdy~\cite{Purdy83} gave an algorithm with average running time $O(n)$
but worst case of order $n^2$. This was improved by
Bojanczyk and Brent~\cite{rpb096}, 
whose algorithm has worst case running time $O(n)$.
\medskip

Let $g = \GCD(u,v)$, $u' = u/g$, $v' = v/g$.
In~\cite[\S4]{rpb096} an algorithm is given for
reducing the fraction $u/v$ to $u'/v'$ without performing
any divisions (except by powers of two).

\subsection{Parallel Variants and the Class NC}

There is a systolic array variant of
the binary GCD algorithm (Brent and Kung~\cite{rpb077}). 
This takes time $O(\log N)$
using $O(\log N)$ 1-bit processors. 
The overall bit-complexity
is $O((\log N)^2)$.
\medskip

For $n$-bit numbers the systolic
algorithm gives time $O(n)$ using $O(n)$ processors. This is close
to the best known parallel time bound
(Borodin {\em et al}~\cite{Borodin82}).

It is not known if GCD is in the class {\bf NC}.\footnote{
{\bf NC} is the class of problems which can be solved in parallel
in time bounded by a polynomial in $\log L$, where $L$ is the length
of the input, using a number of processors bounded by
a polynomial in $L$. Note that for the GCD problem
$L = \Theta(n) = \Theta(\log N)$,
so we are asking for a time polynomial in $\log\log N$, not in $\log N$.}
This is an interesting open problem
because the basic arithmetic operations of addition, multiplication,
and division are in {\bf NC} (see Cook~\cite{Cook83,Cook85}).
Thus, the operation of computing GCDs is perhaps the simplest arithmetic
operation which is {\em not} known to be in {\bf NC}.

It is conceivable that testing coprimality,
i.e. answering the question of whether\\
$\GCD(u,v) = 1$, is
``easier'' than computing $\GCD(u,v)$ in general.
There is evidence that testing coprimality
{\em is} in {\bf NC} (Litow~\cite{Litow99}). 

\section{A Heuristic Continuous Model} 
\label{sec:heuristic}

To analyse the expected behaviour of Algorithm~B,
we can follow what Gauss~\cite{Gauss} did for the classical algorithm.
This was first attempted in~\cite{rpb037}.
There is a summary in
Knuth~\cite[\S4.5.2]{Knuth97}. 
\medskip

Assume that the initial inputs $u_0$, $v_0$ to Algorithm~B
are uniformly and independently distributed in $(0, N)$,
apart from the restriction that they are odd.
Let $(u_n, v_n)$ be the value of $(u,v)$ after $n$ iterations
of step B3.
\medskip

Let \[x_n = {\min(u_n,v_n) \over \max(u_n,v_n)}\] and let $F_n(x)$
be the probability distribution function of $x_n$ (in the limit
as $N \to \infty$). Thus $F_0(x) = x$ for $x \in [0,1]$.

\subsection{A Plausible Assumption}

We make the assumption%
\footnote{Vall\'ee does not make this assumption.
Her results are mentioned in \S\S\ref{sec:algV}--\ref{sec:vallee}.
They show that the assumption is correct in the limit as
$N \to \infty$.}
that
$\Val_2(t)$ takes the value $k$ with probability $2^{-k}$ at step B2.
The assumption is plausible because $\Val_2(t)$ at
step B2 depends on the least significant bits
of $u$ and $v$, whereas the comparison at
step B3 depends on the most significant bits,
so one would expect the steps to be (almost) independent
when $N$ is large. In fact, this independence is exploited
in the systolic algorithms~\cite{rpb096,rpb082,rpb077}
where processing elements perform
operations on the least significant bits without waiting for information
about the most significant bits.

\subsection{The Recurrence for $F_n$}

Consider the effect of steps B2 and B3. We can assume that initially $u > v$,
so $t = u - v$. If $\Val_2(t) = k$ then $X = v/u$ is transformed
to
\[
X' = \min\left({{u-v}\over {2^k v}}\;, {{2^k v} \over {u-v}} \right)
   = \min\left({1-X \over 2^k X}\;, {2^k X \over 1-X}\right)\;.
\]
It follows that $X' < x$ iff
\[X < {1 \over 1 + 2^k/x} \;\;{\rm or}\;\; X > {1 \over 1 + 2^k x}\;. \]
Thus, the recurrence for $F_n(x)$ is
\beq
F_{n+1}(x) = 1 +
	\sum_{k\ge 1} 2^{-k} \left(
	F_n\left({1 \over 1 + 2^k/x}\right) -
	F_n\left({1 \over 1 + 2^k x}\right)
	\right)						\label{eq:Ffeq}
\eeq
with initial condition $F_0(x) = x$ for $x \in [0,1]$.

It is convenient to define
\[
\KnuthG_n(x) = 1 - F_n(x)\;.
\]
The recurrence for
$\KnuthG_n(x)$ is
\beq
\KnuthG_{n+1}(x) =
	\sum_{k\ge 1} 2^{-k} \left(
	\KnuthG_n\left({1 \over 1 + 2^k/x}\right) -
	\KnuthG_n\left({1 \over 1 + 2^k x}\right)
	\right)						\label{eq:Gfeq}
\eeq
and $\KnuthG_0(x) = 1-x$ for $x \in [0,1]$.

\subsection{The Recurrence for $f_n$}

Differentiating the recurrence~(\ref{eq:Ffeq})
for $F_n$ we obtain (formally)
a recurrence for the probability density
$f_n(x) = F_n'(x)$: 
\beq
f_{n+1}(x) =
 \sum_{k\ge 1}
 \left(
 \left({1 \over x + 2^k}\right)^2
 f_n\left({x \over x + 2^k}\right) +
 \left({1 \over 1 + 2^k x}\right)^2
 f_n\left({1 \over 1 + 2^k x}\right)
 \right)
\;. 							\label{eq:ffeq}
\eeq
It was noted in~\cite[\S5]{rpb037} that the coefficients in this
recurrence are positive, and that the recurrence preserves the
$L_1$ norm of nonnegative functions (this is to be expected, since
the recurrence maps one probability density to another).

\subsection{Operator Notation}

The recurrence for $f_n$ may be written as
\[
f_{n+1} = \calB_2 f_n,
\]
where the operator $\calB_2$ is the case $s = 2$ of a more general operator
$\calB_s$ which is defined in~(\ref{eq:op3}) of~\S\ref{subsec:operators}.

\section{Conjectured and Empirical Results}
\label{sec:empirical}

In the 1976 paper~\cite{rpb037}
we gave numerical and analytic evidence (but no proof)
that $F_n(x)$ converges to a limiting distribution $F(x)$ as $n \to \infty$,
and that $f_n(x)$ converges to the corresponding probability
density $f(x) = F'(x)$  
(note that $f = \calB_2 f$ so $f$ is a ``fixed point'' of the operator $\calB_2$).
\medskip

Assuming the existence of $F$, it is shown in~\cite{rpb037}
that the expected number of iterations of Algorithm~B is $\sim K\lg N$
as $N \to \infty$, where $K = 0.705\ldots$ is a constant given by
\beq
K = {\ln 2 / E_\infty}\;,				\label{eq:K2a}
\eeq
and\footnote{We have corrected a typo in~\cite[eqn.~(6.3)]{rpb037}.}

\beq
E_\infty = \ln 2 + \int_0^1
    \left(\sum_{k=2}^\infty \left({1 - 2^{-k} \over 1 + (2^k - 1)x}\right) -
	{1 \over 2(1+x)}\right)\;F(x)\;dx\;.		\label{eq:compE}
\eeq

\subsection{A Simplification}

We can simplify the expressions~(\ref{eq:K2a})--(\ref{eq:compE})
for $K$ to obtain
\beq
K = 2/b\;,						\label{eq:K2b}
\eeq
where
\beq
b = 2 - \int_0^1 \lg(1-x) f(x)\;dx\;.			\label{eq:simpler1}
\eeq
Using integration by parts we obtain an equivalent expression
\beq
b = 2 + {1 \over \ln 2}
	\int_0^1 {1 - F(x) \over 1 - x}\;dx\;.		\label{eq:simpler2}
\eeq
For my direct proof of~(\ref{eq:simpler1})--(\ref{eq:simpler2}),
see Knuth~\cite[\S4.5.2]{Knuth97}.
The idea is to consider the expected change in $\lg(uv)$
with each iteration of Algorithm~B
(to obtain the equivalent but more complicated expression~(\ref{eq:compE})
we considered the expected change in $\lg(u+v)$).

\section{Another Formulation~-- Algorithm~V}
\label{sec:algV}

It will be useful to rewrite Algorithm~B in the following
equivalent form (using pseudo-Pascal):

\begin{description}
\item[Algorithm~V] \{ Assume $u \le v$ \}

\pn {\bf while} $ {u } \ne v$  {\bf do}\\
\ia {\bf begin}\\
\ia {\bf while} $ u < v$ {\bf do}\\
\ib   {\bf begin}\\
\ib   $j \leftarrow \Val_2(v-u);$\\
\ib   $v \leftarrow {(v-u)/2^j};$\\
\ib   {\bf end};\\
\ia $u \leftrightarrow v$;\\
\ia {\bf end};\\
\pn {\bf return} $u$.
\end{description}

\subsection{Continued Fractions}

Vall\'ee~\cite{Va-Algo}
shows a connection between Algorithm~V
and continued fractions of a certain form
\[{u \over v} = {1\over a_1 + {\displaystyle 2^{k_1} \over
 \displaystyle a_2 +
{\displaystyle 2^{k_2} \over \displaystyle{\ \ \ \ \ \
 \ddots\;\; 
 +{ 2^{{k_{r-1}^{{\ }^{\ } }}}\over
 \displaystyle a_r + 2^ {k_r} }}}}}
\]
which by convention we write as
\beq
{u \over v} = 1/a_1 + 2^{k_1}/a_2 + 2^{k_2}/\ldots /(a_r + 2^{k_r})\;.
								\label{eq:Vcf}
\eeq
Here $a_j$ is odd, $k_j > 0$, and $0 < a_j < 2^{k_j}$
(excluding the trivial case $u=v=1$).

\subsection{Some Details of Vall\'ee's Results}

Algorithm~V has two nested loops. The outer loop exchanges $u$ and $v$.
Between two exchanges, the inner loop performs a sequence of subtractions
and shifts which can be written as
\begin{eqnarray*}
v   &\rightarrow& u + 2^{b_1}v_1;\\
v_1 &\rightarrow& u + 2^{b_2}v_2;\\
    &\cdots&\\
v_{m-1} &\rightarrow& u + 2^{b_m}v_m
\end{eqnarray*}
with $v_m \le u$.
\medskip

If $x_0 = u/v$ at the beginning of an inner loop, the effect of the inner
loop followed by an exchange is the rational $x_1 = v_m/u$ defined by
\[x_0 = \frac{1}{a+2^kx_1}\;,\]
where $a$ is an odd integer given by
\[a = 1 + 2^{b_1} + 2^{b_1+b_2} + \cdots + 2^{b_1 + \cdots + b_{m-1}}\;,\]
and the exponent $k$ is given by
\[k = b_1 + \cdots + b_m\;.\]
Thus, the rational $u/v$, for $1 \le u < v$,
has a unique {\em binary continued fraction expansion}
of the form~(\ref{eq:Vcf}).
Vall\'ee studies three parameters related to this continued fraction:
\begin{enumerate}
\item The height or the depth (i.e.~the number of exchanges) $r$.
\item The total number of operations necessary to obtain the expansion
(equivalently, the number of times step B2 of Algorithm~B is performed):
if $p(a)$ denotes
the number of ``1''s in the binary expansion of the integer $a$,
it is equal to
$p(a_1) + p(a_2)+ \cdots + p(a_r)$. 
\item The total number of one-bit shifts,
i.e.~the sum of exponents of 2 in the numerators of the binary continued
fraction, $k_1 + \cdots + k_r$.
\end{enumerate}

\subsection{Vall\'ee's Theorems}

Vall\'ee's main results give the average values of the
three parameters above: the average values are asymptotically $A_i \ln N$ for
certain computable constants
$A_1, A_2, A_3$ related to the spectral properties of an operator
$\calV_2$ which is defined in~(\ref{eq:op4}) of~\S\ref{subsec:operators}.
Clearly the constant $K$ of \S\ref{sec:empirical} is $A_2\ln 2$.

\subsection{Some Useful Operators}
\label{subsec:operators}

Operators $\calB_s$, $\calV_s$, $\calU_s$, ${\tcU}_s$,
useful in the analysis of the binary Euclidean algorithm,
are defined by
\beq
\calB_s [f](x) =
 \sum_{k\ge 1}
 \left(
 \left({1 \over x + 2^k}\right)^2
 f\left({x \over x + 2^k}\right) +
 \left({1 \over 1 + 2^k x}\right)^2
 f\left({1 \over 1 + 2^k x}\right)
 \right)\;, 							\label{eq:op0}
\eeq
\beq
\calV_s [f](x) =
\sum_{ k\ge 1} \sum_{ a \; {\rm odd},\atop {0 <  a < 2^k}}
\left({1 \over a+ 2^k x}\right)^{s}
\  f\left({1 \over a+ 2^k x}\right)\;, 				\label{eq:op4}
\eeq
\beq
\calU_s [f](x) =
\sum_{k \ge 1} \left({1 \over 1+ 2^k x}\right)^{s} \,
f \left({1 \over 1+ 2^k x}\right)\;,				\label{eq:op1}
\eeq
\beq
{\tcU}_s [f](x) =
\left({1 \over x}\right)^s  \calU_s [f]\left({1 \over x}\right)\;.
								\label{eq:op2}
\eeq
In these definitions $s$ is a complex variable, and the operators
are called Ruelle operators~\cite{Ruelle}.
They are linear operators acting on certain
function spaces.
It is immediate from the definitions that
\beq
\calB_s = \calU_s + {\tcU}_s,					 \label{eq:op3}
\eeq
The case $s = 2$ is of particular interest.
$\calB_2$ encodes the effect of one iteration of the inner ``while'' loop
of Algorithm~V, and
$\calV_2$ encodes the effect of one iteration of the outer ``while'' loop.
See Vall\'ee~\cite{Va-ANTS,Va-Algo} for further details.

\subsection{History and Notation}

$\calB_2$ (denoted $T$) was introduced in~\cite{rpb037},
and was generalised to $\calB_s$ by Vall\'ee.
$\calV_s$ was introduced by Vall\'ee~\cite{Va-ANTS,Va-Algo}.
We shall call
\begin{itemize}
\item $\calB_s$ (or sometimes just $\calB_2$)
	the {\em binary Euclidean operator}
and
\item $\calV_s$ (or sometimes just $\calV_2$)
	{\em Vall\'ee's operator}.
\end{itemize}

\subsection{Relation Between the Operators}

The binary Euclidean operator and Vall\'ee's operator
are closely related, as
Lemma~\ref{lem:operators} and Theorem~\ref{thm:operators1} show.

\begin{lemma}
\label{lem:operators}
\[
\calV_s = \calV_s {\tcU}_s + \calU_s.					
\]
\end{lemma}
\medskip

\noindent{\bf Proof.} From~(\ref{eq:op4}),
\[\hspace*{-2em}\calV_s [{\tcU}_s [f]](x) =
	\sum_{ k\ge 1} \sum_{ a \; {\rm odd},\atop {0 <  a < 2^k}}
	\left({1 \over a+ 2^k x}\right)^{s}
\  {\tcU}_s [f]\left({1 \over a+ 2^k x}\right)			
\]
but, from~(\ref{eq:op1}) and~(\ref{eq:op2}),
\[
{\tcU}_s [f](y) =
\sum_{m \ge 1} \left({1 \over 2^m + y}\right)^{s} \,
f \left({1 \over 1 + 2^m/y}\right).				
\]
On substituting $y = 1/(a + 2^k x)$ we obtain

\[\calV_s [{\tcU}_s [f]](x) =
 \sum_{ k\ge 1} \sum_{ a \; {\rm odd},\atop {0 <  a < 2^k}}
\sum_{m \ge 1} \left({1 \over 1 + 2^m a+ 2^{k+m} x}\right)^{s}
\; f\left({1 \over 1 + 2^m a+ 2^{k+m} x}\right).	      
\]
Thus, to show that
\[
\calV_s[f](x) = \calU_s [f](x) + \calV_s {\tcU}_s [f](x)      
\]
it suffices to observe that the set of polynomials
\[
\{a' + 2^{k'} x \st k' \ge 1,\; a' \; {\rm odd},\; 0 < a' < 2^{k'}\}
\]
is the disjoint union of the two sets
\[
\{1 + 2^k x \st k \ge 1\}
\]
and
\[
\{1 + 2^m a + 2^{k+m} x \st
  k \ge 1,\; m \ge 1,\; a \; {\rm odd},\; 0 < a < 2^k\}.
\]
To see this, consider the two cases $a' = 1$ and $a' > 1$.
If $2^{k'} > a' > 1$ we can write $a' = 1 + 2^m a$, $k' = k+m$,
for some (unique) odd $a$ and positive $k$,~$m$.			\QED

\subsection{Algorithmic Interpretation}

Algorithm V gives an interpretation of
Lemma~\ref{lem:operators} in the case $s = 2$.
If the input density
of $x = u/v$ is $f(x)$ then execution of the inner
 ``while'' loop followed by the exchange of $u$ and $v$
transforms this density to $\calV_2 [f](x)$. However, by
considering the first iteration of this loop (followed by the exchange
if the loop terminates) we see that the
transformed density is given by
\[
\calV_2 {\tcU}_2 [f](x) + \calU_2 [f](x),
\]
where the first term arises if there is no exchange,
and the second arises if an exchange occurs.

\subsection{Consequence of Lemma \ref{lem:operators}}

The following Theorem gives
a simple relationship between $\calB_s$, $\calV_s$ and $\calU_s$.

\begin{thm}
\label{thm:operators1}
\[
(\calV_s - \calI) \calU_s = \calV_s (\calB_s - \calI)\;.	
\]
\end{thm}
\medskip

\noindent{\bf Proof.} This is immediate from
Lemma \ref{lem:operators} and the definitions of the operators.		\QED

\subsection{Fixed Points}
\label{subsec:fixedp}

It follows immediately from~Theorem~\ref{thm:operators1} that, if
\begin{equation} 
g = \calU_2 f,						\label{eq:op12}
\end{equation}
then
\[
(\calV_2 - \calI) g = \calV_2 (\calB_2 - \calI) f.		
\]
Thus, if $f$ is a fixed point of the operator $\calB_2$,
then $g = \calU_2 f$ is a 
fixed point of the operator $\calV_2$.
(We can not assert the converse without knowing something about
the null space of $\calV_2$.)
{From} a result of Vall\'ee~\cite[Prop.~4]{Va-Algo}
we know that
$\calV_2$, acting on a certain Hardy space
$\calH^2 (\calD)$,
has a unique positive dominant simple eigenvalue $1$,
so $g$ must be (a constant multiple of)
the corresponding eigenfunction (provided $g \in \calH^2 (\calD)$).

\begin{lemma}
\label{lem:f1g1}
If $f$ is a fixed point of $\calB_2$ and
$g$ is given by~{\rm{(\ref{eq:op12})}}, then
\[
f(1) = 2 g(1) = 2 \sum_{k \ge 1}
  \left({1 \over 1 + 2^k}\right)^2
  f\left({1 \over 1 + 2^k}\right)\;.			      
\]
\end{lemma}
\medskip

\noindent{\bf Proof.} This is immediate				
from the definitions of $\calB_2$ and $\calU_2$.		\QED

\section{A Result of Vall\'ee}
\label{sec:vallee}

Using her operator $\calV_s$,
Vall\'ee~\cite{Va-Algo} proved that
\beq
K = {2 \ln 2 \over \pi^2 g(1)}
\sum_{ a \; {\rm odd},\atop a > 0}
2^{-\lfloor \lg a \rfloor} 
G\left({1 \over a}\right) 				\label{eq:proved1}
\eeq
where $g$ is a nonzero fixed point of $\calV_2$
(i.e. $g = \calV_2 g \ne 0$)
and $G(x) = \int_0^x g(t)\,dt\;.$
This is the only expression for~$K$ which has been proved rigorously.
\medskip

Because $\calV_s$ has nice spectral properties,
the existence and uniqueness (up to scaling) of $g$ can be established.

\subsection{A Conjecture of Vall\'ee}

Let
\beq
\lambda = f(1)\;,					\label{eq:lambda}
\eeq 
where $f$ is the limiting probability density
(conjectured to exist) as in~\S\ref{sec:empirical}.
$\lambda$ and $K$ are fundamental constants which are not known to have
simple closed form expressions~-- to evaluate them numerically we seem to
have to approximate a probability density $f(x)$ (or $g(x)$)
or the corresponding distribution $F(x)$ (or $G(x)$).
Vall\'ee (see Knuth~\cite[\S4.5.2(61)]{Knuth97}) conjectured that
\[ 
	{\lambda \over b} = {2 \ln 2 \over \pi^2}\;,	
\]
where $b$ is given by~(\ref{eq:simpler1}) or~(\ref{eq:simpler2}).
Equivalently, from~(\ref{eq:K2b}), her conjecture is that
\beq
K\lambda = {4\ln 2 \over \pi^2} \;.			\label{eq:conj1b}
\eeq
Vall\'ee proved the conjecture
under the assumption that the operator $\calB_s$
satisfies a ``spectral gap'' condition which has not been proved,
but which is plausible because
it is known to be satisfied by $\calV_s$.
Specifically, a sufficient condition is that the operator,
acting on a suitable function space, has a simple positive
dominant eigenvalue $\lambda_1$,
and there is a positive $\epsilon$
such that all other eigenvalues $\lambda_j$
satisfy $|\lambda_j| \le \lambda_1 - \epsilon$.

\section{Some Relations Between Fixed Points}
\label{sec:formal}

In this section we {\em assume}
that $f$ is a fixed point of the operator $\calB_2$,
$g = \calU_2 f$ as in \S\ref{subsec:fixedp} is a fixed point of
the operator $\calV_2$, and both $f$ and $g$ are analytic functions
(not necessarily regular at $x = 0$). Using analyticity we extend
the domains of $f$, $g$ etc to include the positive real axis $(0,+\infty)$.
Let
\[F(x) = \int_0^x f(t)\;dt\]
and
\[G(x) = \int_0^x g(t)\;dt\]
be the corresponding integrals.
By scaling, we can assume that
\[F(1) = 1\]
but, in view of (\ref{eq:op12}), we are not free to scale $g$.
(See (\ref{eq:Ginf}) below.)

{From} the definition (\ref{eq:op1}) of $\calU_s$ and (\ref{eq:op12}), we have
\[
g(x) = \sum_{k=1}^\infty \left(\frac{1}{1 + 2^kx}\right)^2
	f\left(\frac{1}{1+2^kx}\right)\;,
\]
so, integrating with respect to $x$,
\[
G(x) = \sum_{k=1}^\infty 2^{-k}\left(F(1) -
		F\left(\frac{1}{1+2^kx}\right)\right)\;.   
\]
This simplifies to
\begin{equation}
G(x) = \sum_{k=1}^\infty 2^{-k}\KnuthG\left(\frac{1}{1 + 2^kx}\right)\;.
								\label{eq:anG}
\end{equation}
Although our derivation of (\ref{eq:anG}) assumes $x \in [0,1]$,
we can use (\ref{eq:anG}) to give an analytic continuation of $G(x)$.
Allowing $x$ to approach $+\infty$, we see that
there exists
\[\lim_{x \to +\infty}G(x) = G(+\infty)\]
say, and
\begin{equation}
G(+\infty) = 1\;.						\label{eq:Ginf}
\end{equation}
\medskip

We can use the functional equation~(\ref{eq:Gfeq}) to extend the
domain of definition of $\KnuthG(x)$
to the nonnegative real axis $[0, +\infty)$.
It is convenient to work with $\KnuthG(x) = 1 - F(x)$ rather than with $F(x)$
because of the following result.

\begin{lemma}
\label{lem:oddfunction}
\[
\KnuthG(x) = G(1/x) - G(x)
\]
and consequently
\[
\KnuthG(1/x) = -\KnuthG(x)\;.
\]
\end{lemma}
\medskip

\noindent{\bf Proof.}
This is immediate from~(\ref{eq:Gfeq}) and (\ref{eq:anG}).	\QED

\begin{lemma}
\label{lem:Gsum3}
\[
G(x) = \sum_{k=1}^\infty 2^{-k}
	\sum_{ a \; {\rm odd},\atop 0 < a < 2^k}
	\left(G\left(\frac{1}{a}\right) -
	      G\left(\frac{1}{a + 2^kx}\right)
	\right)\;.
\]
\end{lemma}
\medskip

\noindent{\bf Proof.}
Since $g(x)$ is a fixed point of $\calV_2$, we have
\begin{eqnarray*}
g(x) &=& \calV_2[g](x)\\
     &=& \sum_{k=1}^\infty
	\sum_{ a \; {\rm odd},\atop 0 < a < 2^k}
	\left({1 \over a+ 2^k x}\right)^2
	g\left({1 \over a+ 2^k x}\right)\;.
\end{eqnarray*}
Integrating with respect to $x$ and making the change of variable
$u = 1/(a + 2^kx)$ gives
\[
G(x) = \sum_{k=1}^\infty 2^{-k}
	\sum_{ a \; {\rm odd},\atop 0 < a < 2^k}
	\int_{1/(a+2^kx)}^{1/a} g(u)du\;,
\]
and the result follows. 						\QED
\medskip

\begin{lemma}
\label{lem:Gsum4}
\[
\sum_{k=1}^\infty 2^{-k}
	\sum_{ a \; {\rm odd},\atop 0 < a < 2^k}
	G\left(\frac{1}{a}\right) = 1\;.
\]
\end{lemma}
\medskip

\noindent{\bf Proof.}
Let $x \to +\infty$ in Lemma~\ref{lem:Gsum3} and use~(\ref{eq:Ginf}). \QED
\medskip

The sum occurring in the following Lemma is the same as the sum in
(\ref{eq:proved1}). To avoid confusion we repeat that our normalisation
of $G$ is different from Vall\'ee's, but note that
the right side of (\ref{eq:proved1}) is independent of the normalisation
of $G$ because $g(1)$ appears in the denominator.

\begin{lemma}
\label{lem:Gsum2}
\[
\sum_{ a \; {\rm odd},\atop a > 0}
2^{-\lfloor \lg a \rfloor} G\left({1 \over a}\right) = 1\;.
\]
\end{lemma}
\medskip

\noindent{\bf Proof.}
We can write Lemma~\ref{lem:Gsum4} as
\[
\sum_{ a \; {\rm odd},\atop a > 0}
c_a \; G\left(\frac{1}{a}\right) = 1\;,
\]
where
\[
c_a = \sum_{k \ge 1,\atop 2^k > a} 2^{-k}
    = 2^{-\lfloor\lg a\rfloor}\;.
\]
Thus, the sums occurring in Lemma~\ref{lem:Gsum4} and Lemma~\ref{lem:Gsum2}
are identical.								\QED

\begin{thm}
\label{thm:Keq}
Under the assumptions stated at the beginning of this section,
the expressions {\rm{(\ref{eq:proved1})}} and~{\rm{(\ref{eq:conj1b})}}
are equivalent.
\end{thm}
\medskip

\noindent{\bf Proof.} This follows immediately from
Lemmas~\ref{lem:f1g1} and~\ref{lem:Gsum2}.				\QED

\medskip

\noindent{\bf Remarks.} As noted above, Vall\'ee proved~(\ref{eq:conj1b})
under an assumption about the spectrum of $\calB_s$. Our proof of
Theorem~\ref{thm:Keq} is more direct.
We are not able to prove the equivalence
of (\ref{eq:K2b}) and~(\ref{eq:conj1b}),
but (as described in \S\ref{sec:numerical}) it has been
verified numerically to high precision.

\section{Numerical Results}
\label{sec:numerical}

Using an improvement of the ``discretization method''
of~\cite{rpb037},
with Richardson extrapolation (see \S\ref{subsec:ndetails}) and
the equivalent of more than
50 decimal places (50D) working precision, we computed the limiting
probability distribution $F$,
then $K$
(using~(\ref{eq:K2b}) and~(\ref{eq:simpler2})), 
$\lambda = f(1)$, and $K\lambda$.
The results were
\begin{eqnarray*}
K 	  &=& 0.7059712461\; 0191639152\; 9314135852\; 8817666677\\
\lambda   &=& 0.3979226811\; 8831664407\; 6707161142\; 6549823098\\
K\lambda  &=& 0.2809219710\; 9073150563\; 5754397987\; 9880385315
\end{eqnarray*}

\noindent These are believed to be correctly rounded values.
\medskip

The computed value of $K\lambda$ agrees with $4 \ln 2 / \pi^2$
to 40 decimals\footnote{In fact the agreement
is to 44 decimals.}, in complete agreement with
Vall\'ee's conjecture~(\ref{eq:conj1b}).

\subsection{Some Details of the Numerical Computation}
\label{subsec:ndetails}

A consequence of Lemma~\ref{lem:oddfunction} is that
$\KnuthG(e^{-y})$ is an {\em odd} function of $y$. This
fact was exploited in the numerical computations.
By discretising with uniform stepsize $h$ in the variable $y$
we can obtain $K$ with error
$O(h^{2r+2})$ after $r$
Richardson/Romberg~\cite{Golub-Ortega,Henrici64}
extrapolations, because the error
has an asymptotic expansion containing only even powers of $h$.

In fact, we found it better to take a uniform stepsize $h$ in the
variable $z = \sqrt{y}$, i.e.~make the change of variables
\[ x = \exp(-z^2)\]
because this puts more points near $x = 0$ and less points in
the ``tail''. We truncated at a point $z_{max}$ sufficiently large that
$\exp(-z_{max}^2)$ was negligible.

To obtain 40D results it was sufficient to take $z_{max} = 11$,
$h = z_{max}/2^{15}$, $r = 7$, iterate the recurrence for $\KnuthG_n$ 81 times
(using interpolation by polynomials of degree $2r+1$ where necessary)
to obtain
$\KnuthG \approx \KnuthG_{81}$ to $O(h^{16})$ accuracy.
Using the trapezoidal rule,	
we obtained $K$ by numerical quadrature
to $O(h^2)$ accuracy,
and then applied seven Richardson extrapolations (using the results for
stepsize $h, 2h, 2^2h, \cdots, 2^7h$)
to obtain $K$ with error $O(h^{16})$.
Similarly, we approximated $\lambda = F'(1)$ by
\[
\lambda \approx \KnuthG(\exp(-h^2))/h^2
\]
and then used extrapolation.
Only three Richardson extrapolations
were needed to obtain $\lambda$ with error $O(h^{16})$ because
the relevant asymptotic expansion includes only powers of $h^4$.

\subsection{Subdominant eigenvalues}
\label{subsec:subdominant}

In order to estimate the speed of convergence of $f_n$ to $f$
(assuming $f$ exists), we need more information on the spectrum of
$\calB_2$. What can be proved~?

Preliminary numerical results indicate that the sub-dominant eigenvalue(s)
are a complex conjugate pair:
\[\lambda_2 = {\overline \lambda_3} = 0.1735 \pm 0.0884i\;,\]
with $|\lambda_2| = |\lambda_3| = 0.1948$ to~4D.

The appearance of a complex conjugate pair
is interesting because in the classical case it is known that
the eigenvalues are all real, and conjectured that (when ordered
in decreasing absolute value) they alternate in sign~\cite{FV1}.

\subsection{Complexity of approximating $K$}
\label{subsec:complexity}

We have several expressions for $K$ which are conjectured to be
equivalent. Which is best for numerical computation of~$K$~?
Suppose we want to estimate $K$ to $n$-bit accuracy, i.e.~with
error $O(2^{-n})$.

We could iterate the recurrence
\[g_{k+1}(x) = \calV_2[g_k](x)\]
to obtain the principal fixed point $g(x)$ of Vall\'ee's operator $\calV_2$.
However, the sum over odd $a$ in the definition of $\calV_2$
appears to require the summation of exponentially many terms.
Similarly, the sum in~(\ref{eq:proved1}) appears to require
exponentially many terms (unless we can assume that $g$ is
scaled so that Lemma~\ref{lem:Gsum2} applies).

Thus, it seems more efficient numerically to approximate the principal
fixed point $f(x)$ of the binary Euclidean operator $\calB_2$
or the corresponding integral $F(x)$ (or $\KnuthG(x) = 1 - F(x)$),
even though the existence of $f$ or $F$ has not been proved.

It seems likely that $f(x)$ is unbounded in a neighbourhood of $x = 0$,
so it is easier numerically to work with $\KnuthG(x)$.
In the sum~(\ref{eq:Gfeq}), the terms are bounded by $2^{-k}$, so we need
take only $O(n)$ terms to get $n$-bit accuracy. (If we used the
recurrence~(\ref{eq:ffeq}) it would not be so clear how many terms
were required.)

Assuming that $\calB_2$ has a positive dominant simple eigenvalue~1
(as seems very likely), 
convergence of $\KnuthG_k(x)$ to $\KnuthG(x)$ is linear, so $O(n)$ iterations
are required.
We have to tabulate $F(x)$ at a sufficiently dense set of points
that the value at any point can be obtained to sufficient accuracy
by interpolation. If the scheme of \S\ref{subsec:ndetails} is used,
it may be sufficient to take $h = O(1/n)$ and use polynomial
interpolation of degree $O(n/\log n)$.
(Here~\cite[ex.~4.5.2.25]{Knuth97} may be relevant.)

The final step, of estimating the integral~(\ref{eq:simpler2})
and $\lambda = f(1)$,
can be done as in \S\ref{subsec:ndetails} with a relatively
small amount of work.
Alternatively, we can avoid the computation of an integral by
using~(\ref{eq:conj1b}). However, the independent computation of $K$
and $\lambda$ provides a good check on the numerical results, since
it is unlikely that any errors in the computed values of
$K$ and/or $\lambda$ would be correlated in such a way as to
leave the product $K\lambda$ unchanged.

Overall, the work required to obtain an $n$-bit approximation to $K$
appears to be bounded by a low-degree polynomial in $n$.
Probably $O(n^4)$ bit operations are sufficient.
It would be interesting to know if significantly faster algorithms
exist. For example, is it possible to avoid the computation
of $\KnuthG(x)$ or a similar function at a large number of points~?

\section{Correcting an Error} 
\label{sec:correction}

In~\cite{rpb037} it was claimed that, for all $n \ge 0$ and $x \in (0,1]$,
\beq
F_n(x) = \alpha_n(x)\lg(x) + \beta_n(x)\;,		\label{eq:claim}
\eeq
where $\alpha_n(x)$ and $\beta_n(x)$ are analytic and regular
in the disk $|x| < 1$. However, {\em this is incorrect},
even in the case $n = 1$.
\medskip

The error appeared to go unnoticed until 1997, when Knuth was
revising Volume~2 in preparation for publication of the third edition.
Knuth computed the constant $K$ using recurrences for the analytic
functions $\alpha_n(x)$ and $\beta_n(x)$, and I computed $K$ directly
using the defining integral and recurrences for $F_n(x)$.
Our computations disagreed in the 14th decimal place~!
Knuth found
\pagebreak[3]
{\nobreak
\[
K = 0.70597~12461~019{\underbar{45~99986}} \cdots
\]
but I found
\[
K = 0.70597~12461~019{\underbar{16~39152}} \cdots
\]
}

We soon discovered the source of the error.
It was found independently,
and at the same time, 
by Flajolet and Vall\'ee.
\medskip

The source of the error is illustrated by~\cite[Lemma~3.1]{rpb037},
which is incorrect, and corrected in
~\cite[solution to ex.~4.5.2.29]{Knuth97}.
In order to explain the error, we need to consider
Mellin transforms (a very useful tool in average-case
analysis~\cite{FS}).

\subsection{Mellin Transforms and Mellin Inversion}

The {\em Mellin transform} of a function\footnote{The functions $f$
and $g$ here are not necessarily related to those occurring in
other sections.}
$g(x)$ is defined by
\[ g^{*}(s) = \int_0^\infty g(x)x^{s-1}dx\;. \]
It is easy to see that, if
\[f(x) = \sum_{k \ge 1} 2^{-k}g(2^k x)\;,\]
then the Mellin transform of $f$ is
\[ f^{*}(s) = \sum_{k \ge 1} 2^{-k(s+1)}g^{*}(s)
	= {g^{*}(s) \over 2^{s+1} - 1}\;. \]
Under suitable conditions we can apply the Mellin inversion formula to obtain

\[ f(x) = {1 \over 2\pi i}\int_{c-i\infty}^{c+i\infty} f^{*}(s)x^{-s}ds\;. \]
\medskip

Applying these results to
\[
g(x) = 1/(1+x)\;,
\]
whose Mellin transform is
\[
g^{*}(s) = \pi/\sin \pi s \;\;{\rm when}\;\; 0 < {\cal R}s < 1\;,
\]
we find
\beq
 f(x) = \sum_{k \ge 1} {2^{-k} \over 1+2^k x}			\label{eq:fD1}
\eeq
as a sum of residues of
\beq
\left({\pi \over \sin \pi s}\right){x^{-s} \over 2^{s+1} - 1} \label{eq:mex}
\eeq
for ${\cal R}s \le 0$.  This gives
\beq
f(x) = 1 + x\lg x + {x \over 2} + xP(\lg x) - {2 \over 1}x^2
	+ {4 \over 3}x^3 - \cdots\;,		\label{eq:correctf}
\eeq
where
\beq
P(t) = {2\pi \over \ln 2}\sum_{n=1}^\infty {\sin 2n\pi t \over
	\sinh(2n\pi^2/\ln 2)}\;.		\label{eq:Pt}
\eeq

\subsection{The ``Wobbles'' Caused by $P(t)$}

$P(t)$ is a very small periodic function:
\[|P(t)| < 7.8 \times 10^{-12}\]
for real $t$.
In~\cite[Lemma~3.1]{rpb037},
the term $xP(\lg x)$ in (\ref{eq:correctf}) was omitted.
Essentially, the poles of (\ref{eq:mex}) off the real axis at
\[s = -1 \pm \frac{2\pi i n}{\ln 2}\;, \;\;\;\; n = 1, 2, \ldots\]
were ignored.%
\footnote{In fact, the incorrect result was obtained without using Mellin
transforms. If I had used them I probably would have obtained the correct
result!}
\medskip

Because of the $\sinh$ term
in the denominator of (\ref{eq:Pt}),
the residues at the non-real poles are tiny, and
numerical computations performed using single-precision
floating-point arithmetic 
did not reveal the error.

\subsection{Details of Corrections}

The function $f(x)$ of~(\ref{eq:fD1}) is called $D_1(x)$ in~\cite{rpb037}.
In (3.29) of~\cite[Lemma~3.1]{rpb037}, the expression for $D_1(x)$
is missing the term $xP(\lg x)$.
\medskip

Equation (3.8) of~\cite{rpb037} is (correctly)
\[F_n(x) = 1 + D_n(1/x) - D_n(x)\]
so in Corollary 3.2 the expression for $F_1(x)$ is
missing a term $-xP(\lg x)$.
\medskip

The statement following Corollary~3.2 of~\cite{rpb037},
that ``In principle we could
obtain $F_2(x), F_3(x)$, etc in the same way as $F_1(x)$'' is dubious
because it is not clear how to handle the terms involving $P(\lg x)$.
\medskip

To quote Gauss [notebook, 1800], who was referring to $F_2(x)$ etc for the
{\em classical} algorithm:
\begin{quotation}
{\em Tam complicat\ae\ evadunt, ut nulla spes superesse videatur.}%
\footnote{They come out so complicated that no hope appears to be left.}
\end{quotation}
\medskip

Corollary 3.3 of~\cite{rpb037},
that $F_{n+1} \ne F_n$, is probably correct, but
the proof given is incorrect because it assumes the incorrect
form~(\ref{eq:claim}) for $F_n(x)$.

\subsection{An Analogy}
\label{subsec:analogy}

Ramanujan 
made a similar error when he gave a formula for
$\pi(x)$ (the number of primes $\le x$)
which essentially ignored the residues of $x^s\zeta'(s)/\zeta(s)$
arising from zeros of $\zeta(s)$ off the real axis.
For further details we refer to
Berndt~\cite{Berndt}, Hardy~\cite{Hardy-Ramanujan},
Riesel~\cite[Ch.~1--3]{Riesel}
and the references given there.

\section{Conclusion and Open Problems}
\label{sec:conc}

Since Vall\'ee's recent work~\cite{Va-ANTS,Va-Algo},
analysis of the average behaviour of the binary Euclidean algorithm
has a rigorous foundation. However, some interesting open questions
remain.
\medskip

For example,
does the binary Euclidean operator $\calB_2$
have a unique positive dominant simple eigenvalue $1$?
Vall\'ee~\cite[Prop.~4]{Va-Algo}
has proved the corresponding result for her operator~$\calV_2$.
Are the various expressions for $K$ given above all provably correct~?
(Only (\ref{eq:proved1}) has been proved.)
Is there an algorithm for the numerical computation of $K$ which is
asymptotically faster than the one described in~\S\ref{subsec:ndetails}~?
How can we give rigorous error bounds on numerical approximations to~$K$~?
\medskip

In order to estimate the speed of convergence of $f_n$ to $f$
(assuming $f$ exists), we need more information on the spectrum of
$\calB_2$. What can be proved~?
As mentioned in~\S\ref{subsec:subdominant},
numerical results indicate that the sub-dominant eigenvalue(s)
are a complex conjugate pair with absolute value about 0.1948.

It would be interesting to compute the spectra of $\calB_2$
and $\calV_2$ numerically, and compare with the classical case,
where the spectrum is real and the eigenvalues appear to
alternate in sign.

In order to give rigorous numerical bounds on the spectra of
$\calB_2$ and $\calV_2$, we need to bound the error caused by
making finite-dimensional approximations to these operators.
This may be easier for $\calV_2$ than for $\calB_2$.

\subsection*{Acknowledgements}

Thanks to:
\begin{itemize}
\item Don Knuth for encouraging me to correct and extend my 1976 results
for the third edition~\cite{Knuth97} of {\em Seminumerical Algorithms}.
Some of the results given here are described in~\cite[\S4.5.2]{Knuth97}.
\item Brigitte Vall\'ee for correspondence and discussions on her
conjectures and results.
\item Philippe Flajolet for his notes~\cite{FS} on Mellin transforms.
\item The British Council for support via an {\em Alliance} grant.
\end{itemize}

An abbreviated version of this report appeared as~\cite{rpb183}.

\vspace*{\fill}
\pagebreak[4]

\end{document}